\documentstyle[12pt,aasms4,psfig]{article}

\lefthead{Makishima et al.}
\righthead{Ultra Luminous X-ray Sources}
\begin{document}

\title{THE NATURE OF ULTRA-LUMINOUS COMPACT\\
X-RAY SOURCES IN NEARBY SPIRAL GALAXIES}
\author{Kazuo {\sc Makishima}, Aya {\sc Kubota}, Tsunefumi {\sc Mizuno}, \\
Tomohisa {\sc Ohnishi}, and Makoto {\sc Tashiro}}  
\affil{Department of Physics, University of Tokyo, 
                7-3-1 Hongo,  Bunkyo-ku, \\  Tokyo 113-0033, Japan}
\author{Yoichi {\sc Aruga}, Kazumi {\sc Asai}, 
Tadayasu {\sc Dotani}, Kazuhisa {\sc Mitsuda},  \\ 
Yoshihiro {\sc Ueda}, Shin'ichiro {\sc Uno}, and  Kazutaka {\sc Yamaoka}}
\affil{Institute of Space and Astronautical Science, 
                3-1-1 Yoshinodai, Sagamihara,\\ Kanagawa 229-8510, Japan}
\author{Ken {\sc Ebisawa}}
\affil{Laboratory of High Energy Astrophysics, NASA/Goddard Space Flight Center,
      Greenbelt, \\ MD 20771, U.S.A.}
\and 
\author{Yoshiki {\sc Kohmura}, and Kyoko {\sc Okada}}
\affil{Harima Institute,
       Institute of Physical and Chemical Research,
       1-1-1 Kouto, Mikazuki-cho, Sayo-gun, Hyogo 679-5148, Japan}

\begin{abstract}
Studies were made of {\it ASCA} spectra of seven
ultra-luminous compact X-ray sources (ULXs) in nearby spiral galaxies; 
M33 X-8 (Takano et al. 1994),
M81 X-6 (Fabbiano 1988b; Kohmura et al. 1994; Uno 1997),
IC~342 Source~1 (Okada et al. 1998),
Dwingeloo~1 X-1 (Reynolds et al. 1997),
NGC~1313 Source~B (Fabbiano \& Trinchieri 1987; Petre et al. 1994),
and two sources in NGC~4565 (Mizuno et al. 1999).
With the 0.5--10 keV luminosities in the range $10^{39-40}$ ergs s$^{-1}$,
they are thought to represent a class of enigmatic 
X-ray sources often found in spiral galaxies.
For some of them, the {\it ASCA} data are newly processed,
or the published spectra are reanalyzed.
For others, the published results are quoted.
The {\it ASCA} spectra of all these seven sources have been described 
successfully with so called multi-color disk blackbody (MCD) emission
arising from optically-thick standard accretion disks around black holes.
Except the case of M33 X-8, the spectra do not exhibit hard tails.
For the source luminosities not to exceed the Eddington limits,
the black holes are inferred to have rather high masses,
up to $\sim 100$ solar masses.
However, the observed innermost disk temperatures of these objects,
$T_{\rm in} = 1.1-1.8$ keV,
are too high to be compatible with the required high black-hole masses,
as long as the standard accretion disks around Schwarzschild black holes are assumed.
Similarly high disk temperatures are also observed from 
two Galactic transients with superluminal motions,
GRO~1655-40 and GRS~1915+105.
The issue of unusually high disk temperature 
may be explained by the black hole rotation, 
which makes the disk get closer to the black hole, and hence hotter. 
\end{abstract}


\section{Introduction}
X-ray radiation from a spiral galaxy
consists of emission from discrete X-ray sources, hot gaseous media,
and in some cases the active galactic nucleus (AGN);
for a review, see Fabbiano (1988b).
Majority of the discrete X-ray sources are 
presumably accreting collapsed objects,
in particular low-mass X-ray binaries (LMXBs; a close binary 
consisting of a weakly-magnetized neutron star and a low-mass star). 
In fact, the X-ray emission from M31, 
with the total 2--10 keV luminosity of $5 \times 10^{39}$ ergs s$^{-1}$,
is dominated by some 100 LMXBs
(Van Speybroeck et al. 1979; Fabbiano, Trinchieri \& van Speybroeck 1987; 
Makishima et al. 1989).
The most luminous ones in M31 have X-ray luminosities 
close to the Eddington limit for a 1.4 $M_\odot$ neutron star,
$L_{\rm E}^{\rm NS} = 2 \times 10^{38} $ ergs s$^{-1}$ (Supper et al. 1997).
Similarly, discrete X-ray sources in M33 have 
X-ray luminosities below $L_{\rm E}^{\rm NS}$ (Long et al. 1996),
except the one (called X-8) to be discussed later. 

In many other nearby spiral galaxies, however, 
there reside point-like off-center X-ray sources 
of which the X-ray luminosity significantly exceeds $L_{\rm E}^{\rm NS}$ 
(e.g., Long 1982; Fabbiano 1988b; Marston et al. 1995; Colbert \& Mushotzky 1999).
Actually, the X-ray luminosity function of point-like non-AGN sources in nearby 
spiral galaxies derived with {\it ROSAT} extends well beyond $L_{\rm E}^{\rm NS}$ 
and reaches even $\sim 2 \times 10^{40}$ ergs s$^{-1}$ (Read et al. 1997). 
Although some of these luminous X-ray objects 
are identified with young supernova remnants 
(e.g., Fabbiano \& Trinchieri 1987; Schlegel 1994),
others have not been identified securely in other wavelengths.
These luminous non-AGN and non-supernova-remnant X-ray objects may be called 
``ultra-luminous compact X-ray sources'' (hereafter abbreviated as ULXs).
The ULXs cannot generally be explained as collections of 
multiple sources with luminosity below $L_{\rm E}^{\rm NS}$ each, 
since many of the ULXs exhibit significant time variability.
Statistical calculations show that these objects are not likely to be 
less luminous foreground objects or background AGNs, either (Fabbiano 1988b).

Assuming that ULXs are single compact objects powered by mass accretion,
there are basically two alternative ways of explaining them.
One is to regard a ULX as a sub-Eddington binary involving a 
relatively massive black hole (BH) with mass up to $\sim 100~M_\odot$,
because the Eddington limit for a body of mass $M$ is given,
when electron scattering dominates the opacity, as
\begin{equation}
L_{\rm E} = 1.5 \times 10^{38} (M/M_\odot) ~~ {\rm ergs~s^{-1}}~~.
\label{eq:L_Ed}
\end{equation}
Here we assume a spherical symmetry,
and the hydrogen to helium composition ratio of 0.76:0.23 by weight.
The BH interpretation may be supported 
by a signature of ULXs as a young population;
they are more abundant in spirals with later morphological types,
and they often locate in spiral arms or HII regions (Fabbiano 1988b).
However, there is not necessarily a consensus 
on the presence of $100~ M_\odot$ BHs.

The other is to regard a ULX as a binary
where the X-ray emission is highly collimated toward us,
or a super-Eddington radiation is maintained.
Then, the object can be accreting BHs in an ordinary mass range, or even LMXBs. 
This interpretation has some empirical support,
because LMC~X-2, an LMXB in the Large Magellanic Cloud, 
exhibits a marginally super-Eddington luminosity 
of $2 \times 10^{38}$ ergs s$^{-1}$ (Tanaka 1997).
However, the difficulty of this alternative is
that there is no known mechanism of producing such an efficient
X-ray beaming, or a steady super-Eddington radiation.

Thus, the nature of ULXs has remained a big 
mystery in the modern X-ray astrophysics.
It affects our basic understanding 
of X-ray properties of normal spiral galaxies,
because a single ULX could overwhelm the summed contribution 
from the entire ordinary LMXBs in the galaxy.
Furthermore, a ULX located near the galaxy center
would (and actually does; Mizuno et al. 1999; Colbert \& Mushotzky 1999) 
mimic a low-luminosity AGN.

Clearly, a key to the issue is provided by detailed X-ray spectroscopy,
which has been enabled by {\it ASCA} (Tanaka, Inoue, \& Holt 1994).
In \S~2, we briefly summarize the ``multi-color disk'' (MCD) modeling of 
emission from optically-thick accretion disks around BHs.
We show in \S~3
that this particular spectral model gives adequate description 
of the {\it ASCA} spectra of a fair number of ULXs,
on condition that the disk temperature is made rather high up to $\sim 2$ keV.
We show in \S~4 that the high disk temperature is also observed 
from Galactic BHB binaries (BHBs) with superluminal jets.
In \S~5, we point out a particular problem
associated with the BH interpretation of ULXs; 
that they exhibit disk temperatures 
which are too high for the large BH masses 
required by their high X-ray luminosities.
We present various attempts to solve the problem,
finally arriving at a hypothesis of spinning stellar-mass BHs.
Throughout the paper, 
errors and uncertainties refer to 90\% confidence limits
unless otherwise stated.

\section{The Multicolor-Disk (MCD) Formalism of Accretion Disk Emission}

\subsection{Assumed spectral model}
The X-ray emission from a BHB, in so called soft (or high) state,
consists of a bright soft component (White \& Marshall 1984) 
with a characteristic temperature of $\sim 1$ keV,
and a hard tail.
The soft component is thought to originate from 
an optically thick accretion disk around the BH, 
of which the standard accretion disk model 
by Shakura \& Sunyaev (1973) provides a prototypical formalism.
There are in fact many recent progresses in the theory of accretion disks,
including the concept of slim accretion disk (Abramowicz et al. 1988)
and advection-dominated accretion flow (ADAF; Narayan \& Yi 1995).
A unified picture is presented by by Esin, McClintock, \& Narayan (1997).
Nevertheless, as long as the high (soft) state of BHBs are concerned,
there is no strong observational evidences
for deviations from the standard Shakura \& Sunyaev type accretion disk.
Therefore, we take the standard accretion disk as a start point,
and employ its simple mathematical approximation called 
multi-color disk blackbody (MCD) model (Mitsuda et al. 1984).
The model is known to give a successful physical description of the 
X-ray spectra of many BHBs (Makishima et al. 1986; Ebisawa et al. 1993;
Mineshige et al. 1994;  Tanaka \& Lewin 1995; Tanaka \& Shibazaki 1996; 
Zhang, Cui \& Chen 1997;  Dotani et al. 1997; Kubota et al. 1998).

The MCD model is a superposition of many blackbody elements,
up to a certain maximum color temperature 
(sometimes simply called dsik temperature) $T_{\rm in}$
that is expected to occur near the innermost disk boundary.
The local disk temperature $T(R)$ is considered to scale as 
\begin{equation}
T(R) \propto R^{-3/4}~~,
\label{eq:gradT}
\end{equation}
where $R$ is radial distance from the BH.
Although an exact treatment of the disk emission must take into 
account the inner disk boundary condition and relativistic effects
(e.g. Ebisawa, Mitsuda \& Hanawa 1991), 
the MCD formalism is known to give a reasonable
approximation (e.g., Ebisawa 1991; Dotani et al. 1997; Kubota et al. 1998),
providing a simple physical insight 
and a straightforward comparison among different BHs. 
In particular, the validity of equation (\ref{eq:gradT}) has been confirmed
observationally in the case of Nova Muscae (Mineshige et al. 1994).

\subsection{Assumed geometry}
Assuming a flat disk geometry with inclination $i$ and distance $D$,
the bolometric luminosity of an optically thick accretion disk is given as
\begin{equation}
L_{\rm bol} = 2 \pi D^2 \cdot f_{\rm bol} \; (\cos i)^{-1}~~,
\label{eq:f_bol}
\end{equation}
where $f_{\rm bol}$ is the bolometric flux calculated from the observed 
band-limited flux, via bolometric correction using the MCD model.
This $L_{\rm bol}$ is related to the maximum disk color temperature $T_{\rm in}$ 
and the innermost disk radius $R_{\rm in}$ as
\begin{equation}
L_{\rm bol} = 4 \pi( R_{\rm in}/\xi)^2 \sigma (T_{\rm in}/\kappa)^4~~.
\label{eq:L_bol}
\end{equation}
Here $\sigma$ is the Stefan-Boltzmann constant,
$\kappa \sim 1.7$ (e.g. Shimura \& Takahara 1995) is 
ratio of the color temperature to the effective temperature,
or ``spectral hardening factor'',
and $\xi$ is a correction factor reflecting the fact 
that $T_{\rm in}$ occurs at a radius somewhat larger than $R_{\rm in}$.
Kubota et al. (1998) give $\xi = (3/7)^{1/2}(6/7)^3 = 0.412$.
Because $T_{\rm in}$ is directly observable 
and $L_{\rm bol}$ can be estimated through equation (\ref{eq:f_bol}),
we can solve equation (\ref{eq:L_bol})  as
\begin{equation}
R_{\rm in} = \xi \kappa^2 \sqrt{\frac{L_{\rm bol}}{4 \pi \sigma T_{\rm in}^4}}
           = \xi \kappa^2 \cdot \frac{D}{\sqrt{\cos i}}  
                           \cdot \sqrt{\frac{f_{\rm bol}}{2 \sigma T_{\rm in}^4}}~~,
\label{eq:R_in}
\end{equation}
which provides us with the basic tool to estimate the innermost disk radius $R_{\rm in}$.

When $R_{\rm in}$ is thus estimated, 
we may identify it with the radius 
of the last stable Keplerian orbit
(e.g. Makishima et al. 1986; Tanaka \& Shibazaki 1996).
For a non-spinning BH of mass $M$, 
this condition takes place at $3 \; R_{\rm S}$,
where 
\begin{equation}
 R_{\rm S} = \frac{2\ G M}{c^2} = 2.95 \left( \frac{M}{M_\odot} \right)~~{\rm km}
\label{eq:R_Sch}
\end{equation}
is the Schwarzschild radius, 
$G$ is the gravitational constant, and $c$ is the light speed.
Thus, we may in general write as
\begin{equation}
 R_{\rm in} = 3 \alpha R_{\rm S} = 8.86 \alpha \left( \frac{M}{M_\odot} \right)
~{\rm km} 
\label{eq:RsRin}
\end{equation}
using a positive parameter $\alpha$,
with $\alpha = 1$ corresponding to the Schwarzschild BH.
Conversely, we may obtain an ``X-ray estimated''  BH mass as 
\begin{equation}
 M \equiv M_{\rm XR} = R_{\rm in}/8.86 \alpha ~~~M_\odot~~.
\label{eq:M_disk}
\end{equation}

\subsection{Temperature-luminosity relations}
By substituting equation (\ref{eq:RsRin}) into equation (\ref{eq:L_bol}),
we obtain a temperature-luminosity relation 
for an accretion disk around a BH of mass $M$, as
\begin{equation}
L_{\rm bol} = 7.2 \times 10^{38}
              \left( \frac{\xi}{0.41} \right)^{-2}  
              \left( \frac{\kappa}{1.7} \right)^{-4}  \alpha^2 
             \left( \frac{M}{10~M_\odot} \right)^2 
             \left( \frac{T_{\rm in}}{{\rm keV}} \right)^4 ~~{\rm erg~s^{-1}}~.
\label{eq:Aya1}
\end{equation}
Note that the mass accretion rate is unspecified in this relation.

It may be convenient here to write $L_{\rm bol}$ as 
\begin{equation}
L_{\rm bol} = \eta L_{\rm E} \propto \eta M ~,
\label{eq:eta}
\end{equation}
where a non-dimensional parameter $\eta$ means the disk bolometric luminosity
normalized to the Eddington luminosity of equation (\ref{eq:L_Ed}).
Using this and equation (\ref{eq:L_Ed}),
we can then eliminate $M$ from equation (\ref{eq:Aya1}),
to obtain another form of temperature-luminosity relation as
\begin{equation}
L_{\rm bol} = 3.1 \times 10^{39}  
             \left( \frac{\xi}{0.41} \right)^{2}
            \left( \frac{\kappa}{1.7} \right)^{4}\alpha^{-2} \eta^2 
             \left( \frac{T_{\rm in}}{{\rm keV}} \right)^{-4}~~{\rm erg~s^{-1}}~.
\label{eq:Aya2}
\end{equation}
Note that the BH mass is unspecified in this relation.

By equating equations (\ref{eq:Aya1}) and (\ref{eq:Aya2}), 
and solving it for $T_{\rm in}$, we obtain an important scaling of 
\begin{equation}
T_{\rm in} = 1.2 \left( \frac{\xi}{0.41} \right)^{1/2} 
        \left( \frac{\kappa}{1.7} \right) \alpha^{-1/2} 
       \eta^{1/4} \left(\frac{M}{10~M_\odot} \right)^{-1/4} ~~~ {\rm keV}.
\label{eq:T_in}
\end{equation}
Thus, a heavier BH tends to show a lower disk temperature.

\section{Analysis of the ULX Spectra}

Although {\it ASCA} has a rather limited angular resolution,
we can utilize its wide-band (0.5--10 keV) spectroscopic capability for those ULXs 
which are relatively isolated from other X-ray sources in the target galaxies,
particularly when helped with the {\it ROSAT} and/or {\it Einstein} images.
The ULXs so far studied with {\it ASCA} include;
the central source (X-8) in M33 (Takano et al. 1994),
so called Source~1 in IC~342 (Okada et al. 1998),
source X-6 in M81 (Kohmura 1994; Uno 1997), 
two sources (A and B) in NGC~1313 (Petre et al. 1994),
one (called X-1) in the newly discovered 
spiral galaxy Dwingeloo~1 (Reynolds et al. 1997),
and two ULXs in NGC~4565 (Mizuno et al. 1999).
Below we review these results, and in some cases, 
reanalyze the published spectra or newly process the {\it ASCA} data.
In Table~\ref{tbl:ULXsummary}, we give a brief summary of these results.

\placetable{tbl:ULXsummary}

\subsection{M33 X-8}

The source X-8 (Long et al. 1981; Gottwald, Pietsch \& Hasinger 1987; 
Trinchieri, Fabbiano \& Peres 1988) at the center of M33
has a 2--10 keV luminosity reaching $1 \times 10^{39}$ ergs s$^{-1}$, 
which makes it the most luminous X-ray source in the Local Group.
In spite of the positional coincidence with the optical nucleus,
X-8 is unlikely to be a low-lumiosity AGN,
because of the 106 day periodicity found 
in its {\it ROSAT} lightcurve (Dubus et al. 1997), 
the lack of nuclear activity in M33 in other wavelengths,
and the stellar kinematic evidence for the absence of a massive BH (Lauer et al. 1998).
Then, considering its luminosity exceeding $L_{\rm E}^{\rm NS}$,
X-8 may be interpreted alternatively as a ULX.

Takano et al. (1994), and subsequently Mizuno (2000),
analyzed the spectra of M33 X-8, 
taken with the Gas Imaging Spectrometer 
(GIS; Ohashi et al. 1996; Makishima et al. 1996)
and the Solid-State Imaging Spectrometer 
(SIS; Burke et al. 1994; Yamashita et al. 1997) onboard {\it ASCA}.
The spectra have been described successfully
with an MCD model of $T_{\rm in} \sim 1.15$ keV,
as quoted in Table~\ref{tbl:ULXsummary},
on condition that an additional power-law component is included
to describe a hard tail above $\sim 5$ keV.
Although the MCD emission is associated with either LMXBs or BHBs,
X-8 is too luminous for an ordinary LMXB,
and its  power-law hard tail is characteristic of BHBs.
Takano et al. (1994) and Mizuno (2000)  therefore conclude 
that X-8 is a mass-accreting stellar-mass ($\sim 10~ M_\odot$) BH.
These results have been reconfirmed by Colbert \& Mushotzky (1999).

Quantitatively, if we adopt $i=0^\circ$,
the measured values of $T_{\rm in}=1.15$ keV 
and $L_{\rm bol} = 3.8 \times 10^{38}$ ergs s$^{-1}$ of X-8 
(Table~\ref{tbl:ULXsummary}) can be consistently reproduced 
by substituting $M=5.5~M_\odot$ and $\eta=0.46$
into equations (\ref{eq:Aya1}) and (\ref{eq:Aya2}),  
together with canonical values of $\alpha=1$, $\xi=0.41$, and $\kappa=1.7$.
If instead, e.g., $i=60^\circ$ is adopted, 
$L_{\rm bol}$ is doubled to $7.6 \times 10^{38}$ ergs s$^{-1}$,
and the solution becomes $M \sim 7.8~M_\odot$ and $\eta \sim 0.65$.
These estimates reconfirm the inference made by Takano et al. (1994).
We are then inspired to regard M33 X-8 
as the nearest and prototypical ULX.

\subsection{IC~342 Source~1}

Another case of significant interest is ``Source~1'' 
(Fabbiano \& Trinchieri 1987; Bregman et al. 1993; Makishima 1994) 
located in a spiral arm of the galaxy IC~342.
As reported by Okada et al. (1998),
its spectrum obtained with {\it ASCA} can be expressed 
by an MCD model of maximum color temperature $T_{\rm in} \sim 1.8$ keV,
whereas the power-law and Bremsstrahlung models disagree with the data.
We here reanalyze the same {\it ASCA} data in the same way as Okada et al. (1998).
Detailed study of the time variation is reported elsewhere (Mizuno 2000).

The derived SIS and GIS spectra of Source~1 are 
presented in Figures~\ref{fig:ULXspec}. 
Here and hereafter, all the {\it ASCA} spectra are
presented after background subtraction, 
but without removing the instrumental responses. 
We fitted the SIS/GIS spectra jointly, using three different spectral
models; power-law, thermal Bremsstrahlung, or MCD,
all modified at low energies with photoelectric absorption.
Then, as shown in Table~\ref{tbl:specfits},
only the MCD model among the three has given an acceptable joint fit, 
in agreement with Okada et al. (1998).
Predictions of the best-fit power-law and MCD models are shown respectively
in Figure~\ref{fig:ULXspec}a and Figure~\ref{fig:ULXspec}b as solid histograms. 

The successful MCD fit encourages us to interpret this source
as an accreting BH, like M33 X-8.
However, the BH must be fairly massive (Table \ref{tbl:ULXsummary}) 
for the high source luminosity to stay within the Eddington limit.
More importantly, the measured disk temperature, $T_{\rm in} \sim 1.8$ keV, 
much exceeds those (0.4--1.0 keV) typically found 
among established BHBs in the Milky Way and Magellanic clouds,
in contradiction to the prediction of equation (\ref{eq:T_in})
that more massive BHs should have lower disk temperatures.
These problems, pointed out by 
Okada et al. (1998), Mizuno et al. (1999) and Mizuno (2000),
are found commonly with other ULXs, as revealed hereafter.

We further tried two other spectral models of convex shapes.
One is unsaturated Comptonization (UC) model by Sunyaev \& Titarchuk (1980),
describing thermal electrons of temperature $T_{\rm e}$ 
Compton up-scattering soft seed photons into X-rays.
It has given a fairly good fit to the spectra of IC~342 Source~1,
although not as good as the MCD model (Table~\ref{tbl:specfits}).
This is of no surprise,
because the MCD and UC models can take nearly identical shapes (Ebisawa 1991),
when $T_{\rm e}$ is made close to $T_{\rm in}$ 
and the optical depth for scattering, $\tau_{\rm es}$, is adjusted.
The other is a broken power-law,
often used to approximate synchrotron radiation spectra 
emitted from jet-dominated sources.
This model, modified with photoelectric absorption,
gave a value of chi-square in between those 
from the MCD and Bremsstrahlung fits (Table~\ref{tbl:specfits}).
Physical meanings of these alternative modelings are examined in \S~5.1.

\placefigure{fig:ULXspec}
\placetable{tbl:specfits}

\subsection{M81 X-6}

Because of the explosion of the supernova SN1993J,
the region around the spiral galaxy M81 has been observed repeatedly with {\it ASCA}.
These observations have provided valuable information on 
SN1993J (Kohmura et al. 1994; Kohmura 1994; Uno 1997),
the low-luminosity AGN in M81 (Ishisaki et al. 1996; Iyomoto 1999),
and the previously known ULX called M81 X-6 (Fabbiano 1988a)
which lies close to SN1993J.
A merit of X-6 is the accurate knowledge
of the distance to the host galaxy,
$3.6 \pm 0.3$ Mpc (Freedman et al. 1994).

Figure~\ref{fig:ULXspec}c presents the SIS spectrum of X-6 (plus SN1993J), 
produced by Uno (1997) 
who screened the data through standard procedure, 
accumulated photons falling within $1.'5$ of X-6,
and then subtracted background as well as contamination from the M81 nucleus.
The spectrum co-adds two observations, 
conducted on 1994 April 1 through 2 for 28 ks
and on 1994 October 21 through 22 for 37 ks.
The supernova had faded away significantly by these observations,
and its contribution to the present spectrum can be expressed by 
a steep power law having a photon index of 3.0 and a photon flux density of 
$1.0 \times 10^{-4}$ ergs s$^{-1}$ cm$^{-2}$ keV$^{-1}$ at 1 keV (Uno 1997).
Therefore, the contamination is limited to energies below $\sim 2$ keV.
In the present study of M81 X-6 we utilize this SIS spectrum only,
since the poorer spatial resolution of the GIS
increases the supernova contamination.
Further data analysis of X-6, including the intensity-sorted spectroscopy, 
is reported elsewhere (Mizuno 2000).

We fitted the SIS spectrum with five alternative spectral models;
power-law, thermal Bremsstrahlung, MCD, UC, or broken power-law,
all modified at low energies with photoelectric absorption.
The SN1993J contribution was taken into account
as a fixed model component mentioned above.
As summarized in Table~\ref{tbl:specfits},
the results have been qualitatively similar to those from IC~342 Source~1:
the power-law model failed,
the Bremsstrahlung fit is much better but still unacceptable,
while the remaining three models have been generally successful. 
In any case, reservation should be put on the obtained $N_{\rm H}$,
because of the supernova contamination below $\sim 2$ keV.

Based on the MCD fit and equation (\ref{eq:f_bol}), 
the disk bolometric luminosity becomes as given in Table \ref{tbl:ULXsummary};
clearly, X-8 is a ULX, and thanks to the accurate distance,
its luminosity is considerably more reliable than that of IC~342 Source~1.
The BH mass required by the Eddington limit (Table \ref{tbl:ULXsummary})
may be reasonable as long as $i$ is rather small.
However, in reference to equation (\ref{eq:T_in}),
the measured value of $T_{\rm in}$ (Table \ref{tbl:specfits}) 
is too high for the rather high BH mass.

The high value of $T_{\rm in}$ suggests the presence 
of a separate harder spectral component,
e.g., a power-law hard tail normally seen from soft-state BHBs
just as is the case with M33 X-8 (\S~3.1).
Accordingly, we refitted the X-6 spectrum by a sum of an MCD continuum,
and a hard power-law 
of which the photon index is fixed at 2.3, typical of high-state BHBs.
However the data did not require the hard component,
with its 0.5--10 keV flux being at most $18\%$ 
of the total source flux in the same range. 
Furthermore, inclusion of the hard tail component at the allowed upper limit
{\it increased} the MCD temperature slightly, from 1.48 keV to 1.52 keV.
Alternatively, a hard blackbody (BB) component 
of temperature $\sim 2$ keV might be present,
like in X-ray luminous LMXBs (Mitsuda et al. 1984; Tanaka 1997).
Accordingly, we replaced the power-law hard tail
with a blackbody of temperature fixed at 2.0 keV.
However, the data did not require the BB component either,
and its contribution to the overall 0.5--10 keV flux turned out to be at most 14\%.

\subsection{Dwingeloo~1 X-1}

Through {\it ASCA} observations of the recently discovered 
nearby spiral galaxy Dwingeloo~1 located behind the Milky Way,
Reynolds et al. (1997) detected a ULX, and named it X-1.
Its spectrum, presented in Figure~\ref{fig:ULXspec}d,
was described well with several alternative models
because of rather poor data statistics, 
but the MCD fit was not attempted by Reynolds et al. (1997).
We hence reanalyze the same GIS and SIS spectra,
in terms of the MCD model as well as power-law and Bremsstrahlung models.
As shown in Table~\ref{tbl:specfits}, all these models have been acceptable,
with the power-law and Bremsstrahlung results being 
consistent with those reported by Reynolds et al. (1997).
The MCD model is somewhat preferred among the three,
because the absorption associated with it agrees with the line-of-sight Galactic 
value of $N_{\rm H} = 7 \times 10^{21}$ cm$^{-1}$ (see Reynolds et al. 1997); 
the other two models require excess absorption which could be artificial.
The source again exhibits a quite high disk temperature, $\sim 1.8$ keV.

\subsection{NGC~1313 Source~B}

The spiral galaxy NGC~1313 is known to contain three bright X-ray objects;
one identified with SN1978k, 
another called Source~A located near the nucleus,
and the other called Source~B associated with a spiral arm 
(Fabbiano \& Trinchieri 1987).
Here we focus on Source~B
which is regarded as a ULX (Petre et al. 1994; Colbert et al. 1995);
Source~A could be composite.
The NGC~1313 region was observed twice with {\it ASCA},
on 1993 July 12 and 1995 November 29. 

The first {\it ASCA} observation was reported by Petre et al. (1994),
who found that the Source~B spectra can be described well 
with a Bremsstrahlung or a Raymond-Smith model (Raymond \& Smith 1977), 
whereas the power-law model is rejected.
However, they did not try an MCD fit.
We therefore reanalyzed the same SIS and GIS spectra of NGC~1313 Source~B,
which are reproduced in our Figure~\ref{fig:ULXspec}e.
As summarized in Table~\ref{tbl:specfits},
the power-law fit is unacceptable,
and the Bremsstrahlung fit is marginally acceptable;
the derived parameters agree with those of Petre et al. (1994).
The MCD fit is fully acceptable, 
and the obtained disk temperature is again rather high.

We have also analyzed the {\it ASCA} data of Source~B 
from the second {\it ASCA} observation,
by screening the data in the standard way,
and accumulating the SIS and GIS photons separately
over a common region of radius $3'$ centered on the source.
We subtracted the GIS background using blank-sky data,
and the SIS background using source-free regions of the same on-source data.
As shown in Figure~\ref{fig:ULXspec}f and Table~\ref{tbl:specfits},
the MCD fit is again most successful.
As listed in Table~\ref{tbl:specfits},
the source flux decreased by a factor of 2.5 from 1993 to 1995,
meanwhile the disk temperature decreased by a factor of $\sim 1.4$.
There is marginal evidence (Table~\ref{tbl:ULXsummary})
for an increase in the disk radius as the source became fainter.
This phenomenon is reported in further detail by Mizuno (2000).

\subsection{Two ULXs in NGC~4565}

Finally, we refer to the recent results by Mizuno et al. (1999),
who studied X-ray emission from the edge-on spiral galaxy NGC~4565
using {\it ASCA} and {\it ROSAT}.
The galaxy hosts two bright point-like X-ray sources,
the fainter one coincident with the nucleus (hereafter called center source)
while the brighter one displaced $\sim 2$ kpc above the galaxy disk
(hereafter off-center source),
both showing high bolometric luminosities (Table~\ref{tbl:ULXsummary}).
Their {\it ASCA} spectra have been described successfully by 
the MCD model with rather high disk temperatures (Table~\ref{tbl:ULXsummary}),
and marginally well by the Bremsstrahlung model,
but the power-law fit was unacceptable for the off-center source.
Based on the absence of significant spectral absorption,
Mizuno et al. (1999) conclude 
that the center source is in fact a ULX in NGC~4565 rather than an AGN.

\section{Galactic Transients with Super-luminal Jets}

Although no ULXs are known in our Galaxy or M31,
the issue of rather high disk temperature is 
also seen among some Galactic transient sources (Zhang, Cui \& Chen 1997).
The best examples are GRO~J1655$-$40 and GRS~1915+105,
both exhibiting super-luminal radio jets
which have allowed accurate determinations
of the system inclination and distances.
We here analyze the {\it ASCA} data of these two sources,
because their resemblance to ULXs is considered to be meaningful.
As they are too bright for the SIS, we utilize the GIS data only.

\subsection{GRO~J1655$-$40}
The transient source GRO~J1655$-$40 has 
$i=69.^\circ 50 \pm 0.^\circ08$ (Orosz et al. 1997) 
and $D=3.2$ kpc (Hjellming \& Rupen 1995).
It has a secure BH secondary, 
with the optically determined BH mass $M_{\rm opt}$ of
$7.0 \pm 0.2~ M_\odot$ (Orosz et al. 1997)
or $5.5-7.9~ M_\odot$ (Shahbaz et al. 1999).
The {\it RXTE} observations are reported by Mendez et al. (1998).

We reanalyzed the deadtime-corrected {\it ASCA} GIS spectrum of this source,
acquired on 1995 August 15--16
and published by Ueda et al. (1998) as their Figure~2a;
they fitted it by a power-law model with a high-energy cutoff,
to focus upon the iron absorption feature.
We fit it with the MCD model, plus a power-law 
of which the photon index is fixed at 2.36 after Zhang et al. (1997).
The iron absorption feature was represented with a 
spectral notch at 6.95 keV according to Ueda et al. (1998).
We have obtained an acceptable fit with $\chi^2/\nu = 91.6/91$,
in spite of very high signal statistics.
This reconfirms the correctness of the spectral modeling,
and the accuracy of the instrumental calibration (Makishima et al. 1996).
The derived MCD parameters are given in Table \ref{tbl:ULXsummary};
the observed disk bolometric luminosity is 
$\sim 1/8$ of the Eddington limit for $M_{\rm opt}$.
The absorption turned out to be $(9 \pm 1) \times 10^{21}$ cm$^{-2}$.

Through equation (\ref{eq:R_in}), we obtain $R_{\rm in} \sqrt{\cos i} = 11.4$ km,
which in turn yields an X-ray mass estimate of $M_{\rm XR} \sim 2.2~M_\odot$
via equation (\ref{eq:M_disk}).
This is in fact much lower than the $M_{\rm opt}$ quoted above. 
This $M_{\rm XR}$ does not change by more than 10\%
even when forcing the power-law normalization to be zero 
(then the fit becomes unacceptable),
or leaving the power-law slope free.
Thus, the disk temperature is apparently too high,
and the disk radius is too small,
just like in the case of the ULXs described in \S~3.

We also attempted the spectral fitting using
general relativistic accretion disk (GRAD) model 
(Hanawa 1989; Ebisawa, Mitsuda \& Hanawa 1991),
which improves over the MCD formalism by taking into account
the inner boundary condition and relativistic effects.
It has the BH mass $M$ and the accretion rate $\dot{M}$ as free parameters,
with $i$ and $\kappa$ adjustable auxiliary parameters.
We fixed $i$ at $69^\circ.5$ and $\kappa$ at 1.7,
and again fixed the power-law photon index at 2.36.
As shown in Figure~\ref{fig:Jetspec}a,
an acceptable ($\chi^2/\nu = 70.4/91$) 
fit has been obtained, with $M= 2.91 \pm 0.10 ~ M_\odot$
and $\dot{M}= (2.64^{+0.03}_{-0.07}) \times 10^{18}$ g s$^{-1}$.
The value of $N_{\rm H}$ was not much different from that of the MCD fit
(Table~\ref{tbl:ULXsummary}).
Thus, the mass estimate somewhat increases,
but it still remains significantly below $M_{\rm opt}$.

These results reconfirm the report by Zhang, Cui \& Chen (1997),
who used the {\it ASCA} data acquired on the same occasion
to derive $R_{\rm in}  = 22$ km,
employing also $\kappa=1.7$ and relativistic corrections to the MCD results.
This has led Zhang, Cui \& Chen (1997) to propose
that the BH in GRO~J1655$-$40 is rapidly spinning,
and the accretion disk is prograde to the BH rotation.
We come back to the issue in \S~5.6.

\subsection{GRS~1915+105}

Similarly, the transient GRS~1915+105 is a promising BH candidate,
and has been studied by a number of authors 
including Belloni et al. (1997) and Zhang, Cui \& Chen (1997).
It has $i=70^\circ$ and $D=12.5$ kpc (Mirabel \& Rodriguez 1994),
although the optical estimate of the BH mass is not yet available.
Belloni et al. (1997) report $L_{\rm bol} \sim 1.4 \times 10^{39}$ ergs s$^{-1}$,
which indicates $M_{\rm E} \geq 9 M_\odot$.
Then, the disk temperature should be $\leq 1.2$ keV, 
from equation (\ref{eq:T_in}). 
However, in reality, a much higher value of 
$T_{\rm in} = 2.27 \pm 0.03$ keV was obtained with {\it RXTE} 
during the outburst peak (Belloni et al. 1997). 

We analyzed the {\it ASCA} GIS data of GRS~1915+105, 
acquired on 1995 April 20 for 18 ks  
and described partially by Ebisawa (1997) and Kotani et al. (2000).
The GIS data were screened in the standard way,
and the spectra from GIS2 and GIS3 were added together
after appropriate dead-time corrections (Makishima et al. 1996)
as presented in Figure~\ref{fig:Jetspec}b.
Background is completely negligible.
It has been fitted successfully ($\chi^2/\nu=82.5/81$) with the MCD continuum,
on condition that it is multiplied by the following six factors;
(i) a photoelectric absorption by $N_{\rm H} = (4.3 \pm 0.1) \times 10^{22}$ cm$^{-2}$;
(ii) a smeared Fe-K edge with the center energy $7.12 \pm 0.08$ keV,
   optical depth $1.6 \pm 0.2$, and a fixed width of 10 keV;
(iii) a narrow K$_\alpha$ line of He-like and/or H-like iron,
   centered at $6.98 \pm 0.03$ keV with an equivalent width (EW) of $55 \pm 6$ eV,
(iv) a narrow Fe K$_\beta$ line at $8.15 \pm 0.09$ keV with an EW of $20 \pm 6$ eV,
(v) a third narrow absorption line at $1.90 \pm 0.02$  keV with an EW of $19 \pm 2$ eV,
attributable to K$_\alpha$ line of H-like and/or He-like Si;
and 
(vi) a broad absorption line at $2.52 \pm 0.04$ keV, having width of 0.8 keV
and an EW of $22 \pm 3$ eV, presumably a blend of Si-K$_\beta$ and S-K$_\alpha$ lines.
No power-law tail was required.

The derived MCD parameters are given in Table~\ref{tbl:ULXsummary}.
The bolometric luminosity becomes $1.2 \times 10^{39}$ ergs s$^{-1}$,
which is similar to that measured with {\it RXTE}.
The value of $M_{\rm XR}=3.5 \pm 0.4~M_\odot$  (Table~\ref{tbl:ULXsummary})
falls significantly below $M_{\rm E}$.
Thus, as reported by Belloni et al. (1997),
GRS~1915+105 exhibit a very high $T_{\rm in}$ and too small $R_{\rm in}$,
like GRO~J1655$-$40 and the ULXs.

\placefigure{fig:Jetspec}

\section{Discussion}

\subsection{Spectral properties of the sample ULXs}

We have studied the {\it ASCA} spectra of seven ULXs,
partially referring to published results.
The Bremsstrahlung model failed for least in two spectra 
(IC~342 Source~1, and M81 X-6),   
and the power-law fit is generally unacceptable 
except in the faintest objects.
In contrast, all the spectra have been described successfully by the MCD model
(plus a hard power-law in M33 X-8) 
with a relatively high disk temperature in the range 1.0--1.8 keV.
Thus, the seven ULXs form a rather homogeneous sample,
and hence are thought to be the same type of objects.
With a possible exception of Dwingeloo~1 X-1
with rather unconstrained spectral modeling,
none of them are background AGNs,
since no AGN is known to exhibit the MCD-type
spectra in the {\it ASCA} range.

We found that the UC model and the broken power-law model
can also describe fairly well the two ULX spectra of high statistics,
those of IC~342 Source~1 (\S~3.2) and M81 X-6 (\S~3.3).
These models will give acceptable fits to the 
remaining ULX spectra that have poorer statistics.
However if the UC interpretation were correct,
the spectrum should bear a prominent Fe-K edge absorption feature at $\sim 7$ keV,
because the large optical depth for electron scattering
required by the two ULX spectra, $\tau_{\rm es} = 20-30$,
implies a huge hydrogen column density of $5 \times 10^{25}$ cm$^{-2}$,
and because the heavy ions would not be completely ionized 
at the suggested electron temperature of $T_{\rm e} \sim 1$ keV.
Evidently, such a feature is absent in the observed ULX spectra,
making the UC interpretation of the ULX spectra physically unrealistic.
Similarly, the broken power-law model 
(appropriate for synchrotron emission) is physically unrealistic,
since the break in the photon index of 1.1--1.5 required by the two sources
is too abrupt to be interpreted in terms of synchrotron cooling.

These considerations strongly suggest
that the emission from these ULXs actually originates in optically thick 
accretion disks as approximated by the MCD modeling,
and hence the objects are either BHBs or LMXBs.
The disk temperatures of the ULXs are closer 
to those of luminous LMXBs (Mitsuda et al. 1984; Tanaka 1997),
rather than to those of BHBs.
However, in order for LMXBs to appear as ULXs,
the emission must be highly anisotropic, 
and/or super-Eddington by nearly two orders of magnitude.
Furthermore, the hard BB component characterizing the LMXB spectra
is absent, at least in the M81 X-6 spectrum,
with an upper limit of  $14\%$ (\S~3.3).
This is considerably lower than the fractional BB contribution of 
$\sim 50\%$ observed from the most luminous LMXBs (Mitsuda et al. 1984).
We therefore conclude that the ULXs are accreting BHBs
rather than a version of LMXB.
This makes a very important step towards understanding these enigmatic objects.

\subsection{Problems with the accretion-disk interpretation}

In Table~\ref{tbl:ULXsummary},
we compile the disk bolometric luminosity $L_{\rm bol}$ via equation (\ref{eq:f_bol}),
the minimum BH mass $M_{\rm E}$
necessary to make $L_{\rm bol}$ stay within the Eddington limit,
and the disk inner radius $R_{\rm in}$ given by equation (\ref{eq:R_in}).
The assumed distance $D$ to each host galaxy is also given there.
For all ULXs, we have assumed the face-on inclination,
i.e. $i \sim 0$, that makes $M_{\rm E}$ lowest.
This assumption is not necessarily too arbitrary,
because face-on systems must have higher fluxes for a given bolometric
luminosity according to equation (\ref{eq:f_bol}), 
and hence more selectively detectable,
than those with larger inclinations;
we hence retain the assumption of $i \sim 0$ for the ULXs throughout the paper.
When $D$ and $i$ are changed, these quantities scale as
\begin{equation}
M_{\rm E}\propto D^{2} (\cos i)^{-1}, 
~~~M_{\rm XR} \propto R_{\rm in} \propto D \; (\cos i)^{-1/2}~~.
\label{eq:scaling}
\end{equation}

In Table~\ref{tbl:ULXsummary}, $M_{\rm E}$ takes rather high values,
reaching $70-80~M_\odot$ for IC~342 Source~1 and NGC~4565 off-center source.
Although there is not yet a consensus on the formation of BHs in such a mass range,
some (e.g. Fryer 1999) argue that a star heavier than $\sim 40~M_\odot$ 
can directly form a BH without supernova explosion 
(and hence without losing much of the progenitor mass),
and there are a fair number of observations of stars 
in the mass range of $100-150~M_\odot$ (e.g. Krabbe et al. 1995).
Therefore, the high BH mass may not be a big difficulty.

A more profound problem is that, except M33 X-8 (and NGC~1313 Source~B in 1995),
the observed $T_{\rm in}$ is too high for the required high BH mass
(Okada et al. 1998; Mizuno et al. 1999; Mizuno 2000), as mentioned in \S~3.
The high values are unlikely to be an artifact of the data analysis (\S~3),
and substitution of these values into equation (\ref{eq:T_in})
would make the BH masses $<5~ M_\odot$ even at the extreme of $\eta =1$.
This makes self-inconsistent the BH interpretation of ULXs,
where a rather high BH mass is inevitable.
This problem is best visualized by Figure~\ref{fig:Aya_plot},
to be called an X-ray ``H-R diagram'' for BHs,
which summarizes various BHBs and ULXs 
on the plane of $T_{\rm in}$ vs. $L_{\rm bol}$.
There, a family of grid lines with a positive slope 
$(L_{\rm bol} \propto T_{\rm in}^4)$ represent the loci of constant $M$,
expressed by equation (\ref{eq:Aya1}).
The other family of grid lines with a negative slope
$(L_{\rm bol} \propto T_{\rm in}^{-4})$ indicate the loci of constant $\eta$,
expressed by equation (\ref{eq:Aya2}).
In order to consistently interpret an X-ray source as an accreting non-spinning BH,
the data point must fall on the region of 
$\eta < 1$ (non-violation of the Eddington limit) 
and $M > 3~M_\odot$ (heavier than a neutron star) in Figure~\ref{fig:Aya_plot}.
The reality, however, is that six out of the seven ULXs (except M33 X-8) 
fall on the super-Eddington regime in this plot.
As to NGC~1313 Source~B, the 1993 data point is super-Eddington,
whereas the 1995 point is marginally sub-Eddington.

\placefigure{fig:Aya_plot}

The problem may equivalently be stated in the following way.
The observed high values of $T_{\rm in}$, 
together with equation (\ref{eq:R_in}),
yield rather small values of $R_{\rm in}$ (Table~\ref{tbl:ULXsummary}).
Then, the BH mass $M_{\rm XR}$, determined by equation (\ref{eq:M_disk}), 
becomes uncomfortably low.
Quantitatively, the derived $M_{\rm XR}$ falls by a factor of 1.6--6.0
short of the minimum BH mass $M_{\rm E}$ to satisfy the Eddington limit.
The issue may be illustrated in Figure~\ref{fig:MeRin},
to be called a ``mass-radius diagram'' for BHs,
where we compare the values of $R_{\rm in}$ against $M_{\rm E}$ or $M_{\rm opt}$.

It is important to note that the same problem
is found also with the two Galactic jet sources studied in \S~4.
We accordingly plot these two objects as well 
in Figure~\ref{fig:Aya_plot} (more than once for each object)
and Figure~\ref{fig:MeRin}.
Indeed, in the H-R diagram of Figure~\ref{fig:Aya_plot},
GRS~1915+105 falls on the super-Eddington regime,
and GRO~J1655$-$40 is located near the lower-mass boundary of a BH.
Similarly in the mass-radius diagram of Figure~\ref{fig:MeRin}, 
the two jet sources (both referring to our own data)
deviate from the relation of ordinary BHBs (see \S~5.4),
and line up better with the ULXs.

\placefigure{fig:MeRin}

\subsection{Simple solutions}

As first-cut attempts to solve the above problem,
let us consider trivial possibilities, 
namely revisions of the inclination $i$, the Eddington limit, or the distance.

So far, we have assumed $i=0$.
However, as is apparent from equation (\ref{eq:scaling}),
the ratio of $M_{\rm E}/M_{\rm XR}$ increases with $i$
as $\propto (\cos i)^{-1/2}$.
Therefore the problem of high $M_{\rm E} / M_{\rm XR}$ ratio
gets worse if $i$ is increased from $\sim 0$.
Thus, $i$ has already been selected to make the problem least severe. 
Larger values of $i$ would also make the BH mass still higher.

Alternatively, the problems may result from our wrong application of 
the spherically-symmetric Eddington limit of equation (\ref{eq:L_Ed}) 
to the systems considered here,
which are basically axi-symmetric.
However, the critical upper-limit luminosity of an axi-symmetric BH system
does not differ very much from the spherical one 
(e.g., Abramowicz et al. 1988; Beloborodov 1998).
This justifies our use of equation (\ref{eq:L_Ed}).

Yet another simple possibility is 
that the source distances are systematically overestimated:
note $M_{\rm E} / M_{\rm XR} \propto D$ from equation (\ref{eq:scaling}).
However, in order to bring $M_{\rm E}/M_{\rm XR}$ down to unity,
we have to reduce the distances to IC~342, M81, NGC~1313, and NGC~4565
by a factor of 6.0, 2.2, 3.1, and 3.7, respectively.
Such large distance revisions would be unwarranted,
especially for M81 of which the distance is known to within $\pm 10$\%
(Freedman et al. 1994).
Therefore, this attempt is unsuccessful,
 as already noticed by Mizuno et al. (1999) regarding the ULXs in NGC~4565.

\subsection{Possible errors in $\xi \kappa^2$}
As is obvious from equations (\ref{eq:R_in}) and (\ref{eq:M_disk}),
the $M_{\rm E}$ vs. $M_{\rm XR}$ discrepancy may alternatively be caused 
because our choice of $\xi=0.41$ (Kubota et al. 1998) 
and $\kappa=1.7$ (Shimura \& Takahara 1995), or $\xi \kappa^2 =1.18$, is wrong.
In this subsection,  we therefore attempt to ``calibrate'' 
the values of $\xi$ and  $\kappa$ mainly using established BHBs.

The factor $\xi$ corrects the MCD formalism for the 
inner boundary condition of the accretion disk,
in such a way that the disk temperature reaches maximum
at $R_{\rm in}/ \xi$ instead of $R_{\rm in}$.
We may then examine our choice of $\xi=0.41$ (Kubota et al. 1998) 
in reference to the GRAD model mentioned in \S~4.1
that properly takes into account this effect.
We accordingly simulated a number of spectra using the GRAD model,
by changing $M$, $\dot{M}$ and $i$, but fixing $\kappa$ at 1.7.
The spectra were analyzed with the MCD formalism,
employing $\xi =0.41$ and $\kappa=1.7$.
Then, over the inclination range of $0^\circ$ to $75^\circ$
and over $T_{\rm in}= 1.2-2.0$ keV,
the X-ray mass $M_{\rm XR}$ from the MCD analysis agreed within $\pm 25\%$ 
with the BH mass $M$ employed as initial inputs to the GRAD model simulations.

We can also utilize Cygnus X-1, the prime BHB, for this purpose.
Applying the GRAD model to the {\it ASCA} data 
of Cygnus X-1 acquired in the 1997 May soft state, 
and assuming $D=2.5$ kpc and $i=30^\circ$,
Dotani et al. (1997) derived $M_{\rm XR}= 12^{+3}_{-1}~M_\odot$.
If we apply our MCD formalism with $\xi \kappa^2 =1.18$
to the same dataset for Cygnus~X-1,
we obtain $R_{\rm in}=90 \pm 18$ km (Table~\ref{tbl:ULXsummary}), 
and hence $M_{\rm XR}=10 \pm 2 ~M_\odot$,
which is close to the GRAD result by Dotani et al. (1997).
Furthermore, our value of $M_{\rm XR}$ is consistent with
the latest optical estimate of the BH mass in Cygnus X-1,
$M_{\rm opt} = 10.1^{+4.6}_{-5.3} ~M_\odot$ (Herreo et al. 1995).
Thus, our overall knowledge of Cygnus X-1 
supports the value of $\xi \kappa^2 \sim 1.2$. 
We plot Cygnus X-1 in Figures~\ref{fig:Aya_plot} and \ref{fig:MeRin}.

We may further utilize the two Magellanic BHBs, LMC~X-1 and LMC~X-3,
which have well established distances of 50--55 kpc; here we adopt 55 kpc. 
We hence analyzed the {\it ASCA} GIS data of LMC~X-1 and LMC~X-3, 
obtained on 1995 April 2 for 22 ks and 1995 April 14 for 23 ks, respectively.
Their GIS spectra have both been fitted successfully
by a model consisting of an MCD component and a power-law tail,
whenall model parameters are left free to vary.
The derived MCD parameters are given in Table~\ref{tbl:ULXsummary},
while the power-law index turned out to be
$2.36^{+0.14}_{-0.18}$ for LMC X-1 and $2.52^{+0.22}_{-0.25}$ for LMC~X-3.
The photoelectric absorption is 
$N_{\rm H} =(5.9 \pm 0.3) \times 10^{21}$ cm$^{-2}$ in LMC X-1 ,
and $N_{\rm H} < 2.9 \times 10^{21}$ cm$^{-2}$ in LMC~X-3.
The measured disk temperatures, given in Table~\ref{tbl:ULXsummary},
agree with the {\it Ginga} measurements (Ebisawa 1991; Ebisawa et al. 1993). 
Using $i \sim 60^\circ$ for LMC~X-1 (Cowley 1992; Cowley et al. 1995)
and $i = 66^\circ \pm 2^\circ$ for LMC~X-3 (Kuiper et al. 1988),
we obtain the disk radii as given in Table~\ref{tbl:ULXsummary}.
In Figure~\ref{fig:MeRin}, these radii are in good consistency
with the optical mass estimates of 
$M_{\rm opt} = 4-10~M_\odot$ for LMC~X-1 (Cowley 1992; Cowley et al. 1995)
and $M_{\rm opt} = 5.0-7.2~M_\odot$ for LMC~X-3 (Kuiper et al. 1988, $D=55$ kpc).

Thus, the value of $\xi \kappa^2$ we have employed is not particularly wrong,
and may not be changed by more than a factor of 1.5 or so.
This tolerance is not large enough to solve 
the $M_{\rm E}$ vs. $M_{\rm X}$ discrepancy for many of our sources.

\subsection{Luminosity dependences}

Even though we have argued in \S~5.4 
for the justification of our estimates on $R_{\rm in}$,
the disk structure may depend on $L_{\rm bol}$ or $\dot{M}$,
particularly when $L_{\rm bol}$ approaches the Eddington limit.
The electron scattering effect may get severer,
which will increase $\kappa$, e.g., from 1.7 to 1.8--2.0 (Shimura \& Takahara 1995).
Furthermore, the true disk temperature may become higher
than is described by the standard accretion disk model,
because entropy of the accreting matter may start being ``advected'' into the BH  
as described by the slim disk model (Abramowicz et al. 1988).
Recently, Watarai et al. (1999) have shown via calculation
that significant amount of X-rays are radiated even from the region inside 
$3 R_{\rm s}$ in a slim accretion disk accreting close to the Eddington limit.
(We may also refer to Reynolds \& Begelman (1997)
who discuss the effect of accretion flow inside $3 R_{\rm s}$.)
While this mechanism can potentially explain what we have observed,
its effects will vanish when the accretion rate becomes sufficiently low.
Therefore, we are urged to search the available data 
for luminosity-dependent effects, such as apparent changes in $R_{\rm in}$.

As mentioned in \S~5.4.3, LMC~X-3 was observed extensively with {\it Ginga}.
In Figure~\ref{fig:Aya_plot},
we therefore plot all the {\it Ginga} and {\it ASCA} data points for LMC~X-3.
Thus, the luminosity of LMC X-3 varied by a factor of $\sim 7$, 
reaching $\sim 2/3$ of the Eddington limit.
However, the data points are aligned very well with a grid line for a constant mass
as already reported previously (Ebisawa 1991; Ebisawa et al. 1993),
and the implied mass is consistent with $M_{\rm opt}$.
Therefore, we do not see any luminosity-dependent changes in $R_{\rm in}$.

The bright BH transient GS~2000+25, 
discovered with the {\it Ginga} ASM (Tsunemi et al. 1989),
is still more suited for the study of the luminosity dependence of accretion disks,
because of its large luminosity swing.
After the discovery, 
GS~2000+25 was observed with the {\it Ginga} LAC eight times spanning 240 days,
covering almost the outburst peak (Ebisawa 1991).
The obtained 2--30 keV spectra were all expressed well 
with an MCD component plus a hard power-law tail.
In Figure~\ref{fig:Aya_plot},
we plot seven out of the eight {\it Ginga} data points (discarding the last one),
assuming a source distance of 3 kpc and  $i=65^\circ$ after Callanan et al. (1996).
Thus, over a very wide range of $\dot{M}$ up to $\eta \sim 0.7$,
the data points generally distribute along a constant-mass grid line,
as already noticed by Ebisawa (1991).
The value of $M_{\rm X} \sim 7~M_\odot$ indicated by Figure~\ref{fig:Aya_plot}
agrees with $M_{\rm opt}$ measured by various authors
(Filippenko, Matheson \& Barth 1995; Beekman et al. 1996; Callanan et al. 1996),
which cluster around $5-8~M_\odot$.
Thus, the standard-disk interpretation is fully self consistent as to this objects,
with little evidence of luminosity-dependent effects.

The superluminal source GRO~J1655-40 itself
exhibited a large luminosity decline as well.
In Figure~\ref{fig:Aya_plot}, the {\it ASCA} and {\it RXTE} data points
generally align along a constant-mass locus,
even though the absolute value of $M_{\rm XR}$ falls 
considerably below $M_{\rm opt}$ as argued repeatedly.
In short, the small value of $M_{\rm XR}$ 
(or the too high $T_{\rm in}$) of this jet source 
is not specific to its high-luminosity states,
but appears intrinsic to it.

We may further argue against luminosity-dependent effects
by comparing GRO~J1655$-$40 and LMC X-3.
These two objects are very similar in terms of the BH mass, 
inclination, and $L_{\rm bol}$ at the time of 
the {\it ASCA} observation (Table~\ref{tbl:ULXsummary}). 
Nevertheless, their disk temperatures differ systematically by 40\%,
which makes their disk radii significantly different in Figure~\ref{fig:MeRin}.

We therefore conclude that luminosity-dependent changes in the accretion disk,
if any, cannot fully explain the high disk temperature 
seen among a certain class of BHs.
We regard it as an intrinsic property of such BHs.

\subsection{The spinning black hole scenario}

We are finally urged to examine the remaining degree of freedom in our MCD scenario, 
i.e., the assumption of $\alpha=1$ in equation (\ref{eq:RsRin}),
which reflects the fact 
that the stable circular orbit can exist only outside $3\; R_{\rm s}$
in the case of a Schwarzschild (i.e.,  a non-rotating) BH.
If instead the BH is rotating significantly (i.e., a Kerr BH),
and the accretion disk is prograde to it,
the last stable Keplerian orbit gets closer to the BH, 
down to $R_{\rm in}=0.5\; R_{\rm s}$ for a maximally rotating BH.
As a result, we expect a Kerr BH to have a smaller value of $\alpha$,
down to the extreme of $\alpha=1/6$,
than a Schwarzschild BH of the same mass and the same accretion rate.
Then, a Kerr BH can attain a higher value of $T_{\rm in}$,
neglecting for the moment stronger relativistic effects.
In Figure~\ref{fig:MeRin}, the dotted line shows 
the prediction for the maximally rotating BHs,
obtained by substituting $\alpha=1/6$ into equation (\ref{eq:RsRin}).
Thus, except for IC~342 Source~1 for which the distance is rather uncertain,
all the objects now fall on the physically reasonable region.
Therefore we propose that the ULXs contain Kerr BHs.
This idea was first proposed by Zhang, Cui \& Chen (1997), 
to explain the high disk temperatures of GRS~1915+105 and GRO~J1655$-$40,
and subsequently applied by Mizuno et al. (1999) to the two ULXs in NGC~4565.

Of course, investigation of the emission from rotating BHs
must fully take into account relativistic effects.
Such studies have been conduced by several authors,
including, e.g., Asaoka (1987), Zhang, Cui \& Chen (1997), and Beloborodov (1998).
Asaoka (1987) showed 
that the spectrum from an optically-thick accretion disk around a Kerr BH 
is similar in shape to that around a Schwarzschild BH, except a higher temperature.
Zhang, Cui \& Chen (1997) calculated 
how the maximum color temperature of the accretion disk, 
$T_{\rm in}$, depends on the BH rotation.
They show that, when the disk is prograde to a maximally rotating BH,
$T_{\rm in}$ can be higher by a factor of $\sim 3$
than that around a Schwarzschild BH
with the same BH mass and the same mass accretion rate.
This is large enough to account for the high disk temperature of our sample objects.
(This temperature enhancement factor becomes $6^{0.75}=3.8$ 
in terms of the MCD formalism which neglects the general relativistic effects.)

There is another merit in appealing to the BH rotation,
that the radiation efficiency, 
defined as the ratio $L_{\rm bol}/\dot{M}c^2$, 
gets higher as the BH angular momentum increases.
The efficiency is 0.057 for a Schwarzschild BH,
while it increases to $\sim 0.4$ for a maximally rotating BH 
fed with a prograde disk (see Zhang, Cui \& Chen 1997).
Therefore, in a maximally rotating BH,
the same mass accretion rate
can produces 7 times higher luminosity than in a non-rotating BH,
although the same Eddington limit should still apply.

How to make rather massive (several tens $M_\odot$) Kerr BHs is an intriguing issue.
There seems at least three different scenarios.
One is the simplest scenario, to assume
that a massive star with a rapid rotation collapses into a spinning BH.
Another is to invoke a merger of two ordinary stellar-mass BHs,
where the angular momentum is brought in from their orbital motion,
and the BH mass is doubled.
The other is to assume that a slowly spinning BH is
spun up due to angular momentum supplied by the accreting matter,
just as binary X-ray pulsars are spun up.
Among the three scenarios,
the last one is particularly attractive because it can potentially explain 
the large BH mass and the high angular momentum at the same time.
Although  Chen, Cui \& Zhang (1997) argued that this mechanism is inefficient,
the scenario may change significantly 
if the accretion is allowed to become considerably advective
and super critical (Beloborodov 1998).  
More quantitative account of this scenario, however,
is beyond the scope of the present paper.

\section{Summary and Conclusion}

We have shown that the multi-color disk blackbody (MCD) emission model can 
successfully reproduce the 0.5--10 keV {\it ASCA} spectra 
of the seven ULXs in nearby spiral galaxies.
These objects therefore form a homogeneous sample,
and are considered to be mass-accreting BHs
of which the X-ray emission originates from optically-thick accretion disks. 
Their bolometric luminosities,  
reaching $10^{39-40}$ ergs s$^{-1}$ even allowing for possible distance errors,
require the BHs to be relatively massive, up to several tens solar masses.

The innermost disk temperatures of the ULXs
have been found in the range $T_{\rm in} = 1.1-1.8$ keV.
These values significantly exceed those typically found in Galactic and Magellanic BHBs,
and are inconsistent with the high BH mass indicated by the high ULX luminosities.
Similarly high disk temperatures are observed from
the two Galactic superluminal jet sources, GRO~J1655-40 and GRS~1915+105.
This problem cannot be explained away by changing the disk inclination,
the source distance, or the value of $\xi \kappa^2$, within tolerances.
Although an optically-thick advective disk (a slim disk) may
account qualitatively for these observations,
we do not find evidence of luminosity-dependent deviations
from the predictions of the standard disk scinario.

We hence suggest that the BHs in these objects are spinning rapidly (i.e. Kerr BHs),
so that the accretion disks can get closer to the BHs and get hotter.
In short, a ULX may be a Kerr BH with a mass of several tens $M_\odot$,
in which an optically-thick  accretion disk is
radiating at a near-Eddington luminosity.
This provides the first concrete working hypothesis
with which the ULX phenomenon can be investigated.

\vspace{4mm}
We would like to thank Drs. T. Hanawa,  S. Mineshige, 
J. Fukue, and Mr. K. Watarai for helpful discussions.
We also thank members of the {\it ASCA} team.
The present work is supported in part by 
the Grant-in-Aid for Center-of-Excellence, No. 07CE2002, 
from Ministry of Education, Science, Sports and Culture of Japan.

\clearpage

\epsscale{0.8}
\plotone{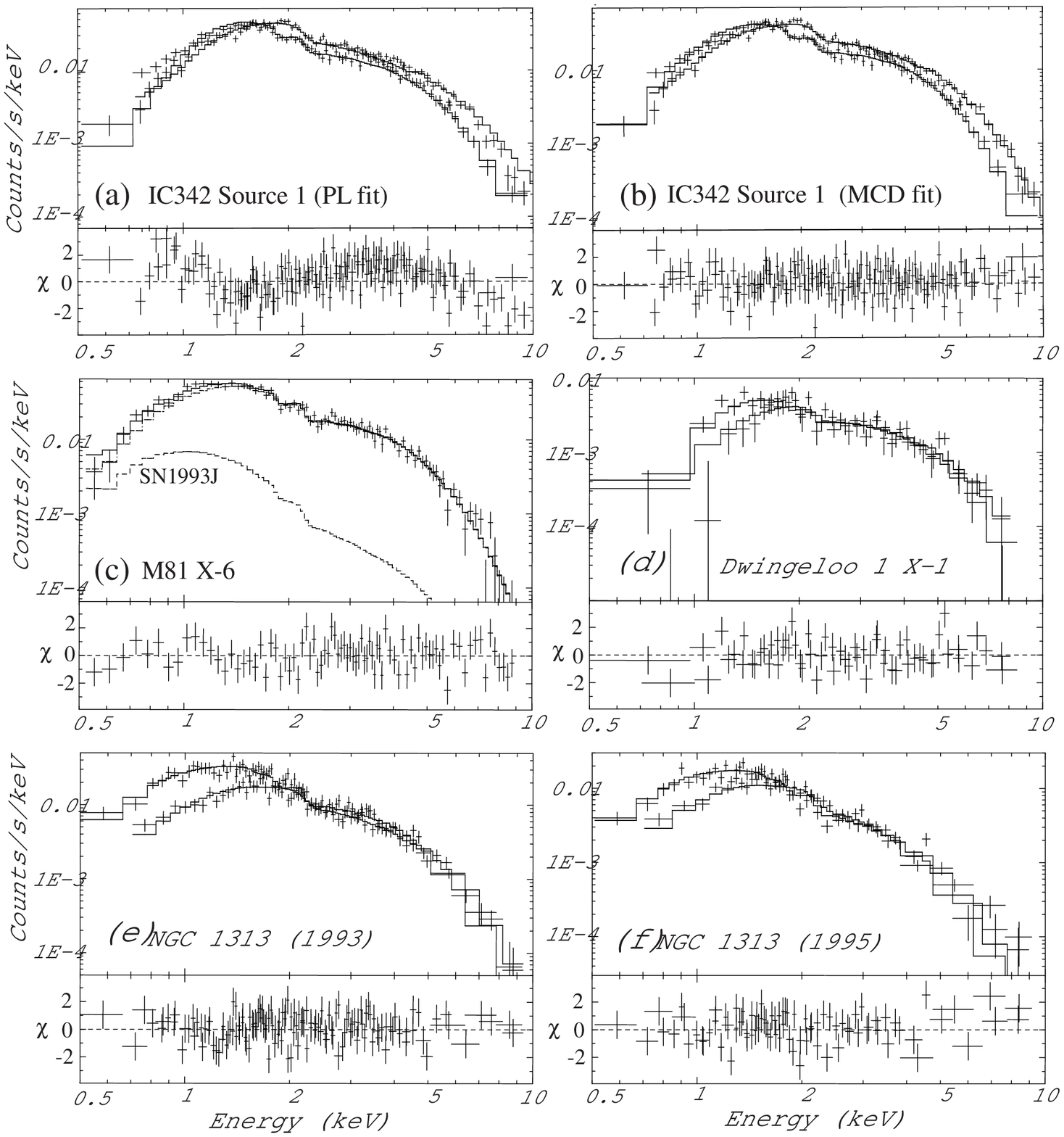}
\figcaption{
The background-subtracted {\it ASCA} spectra of four ULXs, 
shown without removing the instrumental responses.
Except for panel (c), 
the spectra obtained with the SIS (higher counts in lower energies) 
and the GIS (higher counts in higher energies)
are fitted simultaneously with spectral models.
The best-fit model parameters are given in Table~1 and Table~2,
and the fit residuals are presented in bottom of each panel.
(a) Spectra of IC~342 Source 1, fitted with an absorbed power-law.
(b) The same spectra as panel (a), fitted with an absorbed MCD model.
(c) The SIS spectrum of M81 X-6, together with the best-fit absorbed MCD model.
    Contamination from SN~1993J is also indicated.
(d) The Dwingeloo 1 X-1 spectra and the best-fit absorbed MCD model.
(e) Spectra of NGC~1313 Source B obtained in 1993, 
    shown together with the MCD fit.
(f) The same as panel (e), but for the 1995 data.
\label{fig:ULXspec}
}

\clearpage
\epsscale{0.4}
\plotone{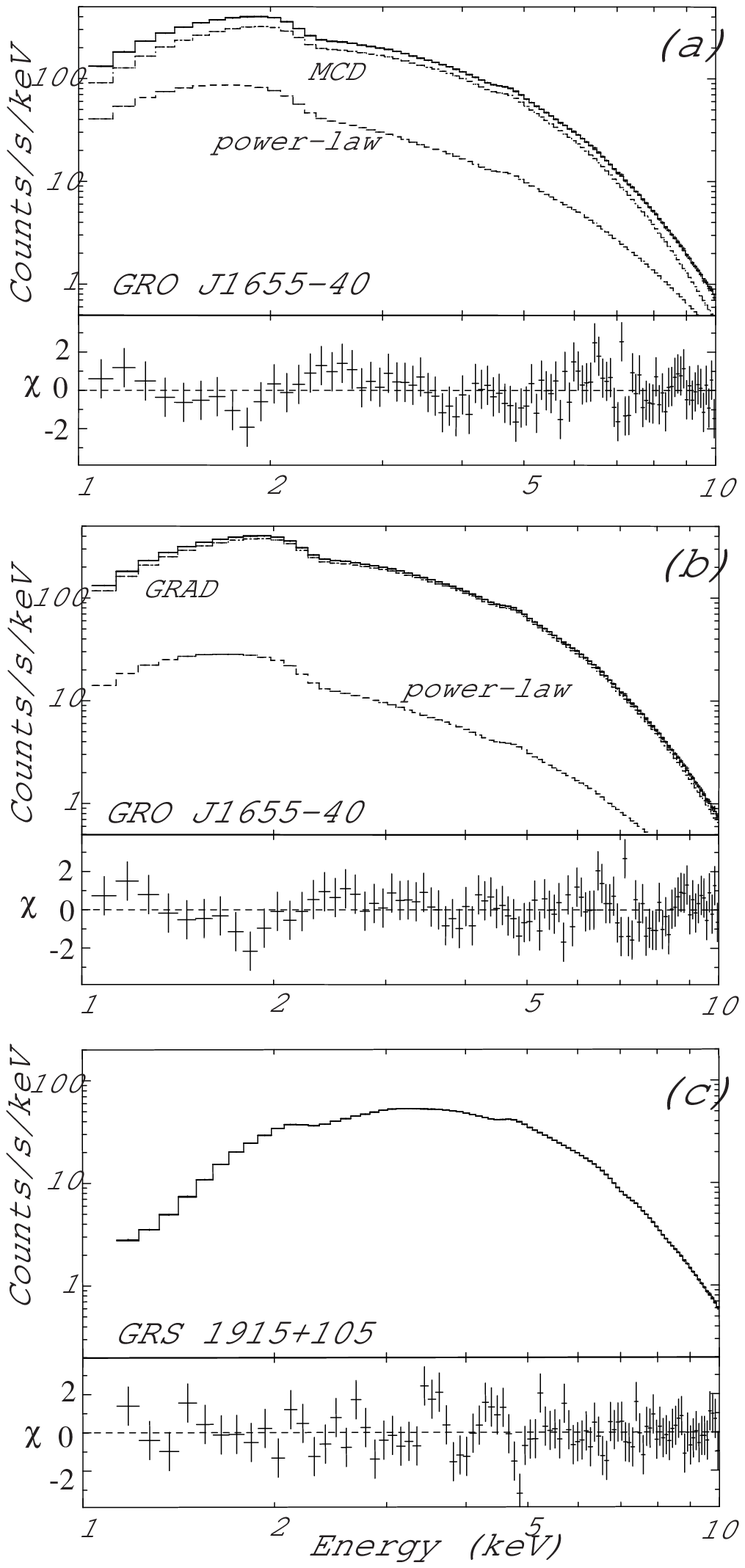}
\figcaption{
 The deadtime-corrected {\it ASCA} GIS spectra 
 of the BH binaries with super-luminal jets,
 fitted with the MCD model or GRAD model.
 The signal counts are so high that the error bars associated with each data point
  are almost invisible.
 The fit parameters are listed in Table~1, and described in \S~4.
(a) Spectrum of GRO~J1655$-$40 obtained on 1995 August 15--16
  (Ueda et al. 1998), fitted with the MCD model plus a power-law.
  The iron-K absorption line is modeled with a notch at 6.95 keV.
(b) The same as panel a, but with the MCD model replaced by the GRAD model.
(c) Spectrum of GRS~1915+105 observed on 1995 April 20, fitted with the MCD model.
  The continuum is modified by photoelectric absorption and smeared iron edge.
  In addition, four  atomic absorption lines are included (see \S~4.2).
\label{fig:Jetspec} 
}

\clearpage
\epsscale{1.0}
\plotone{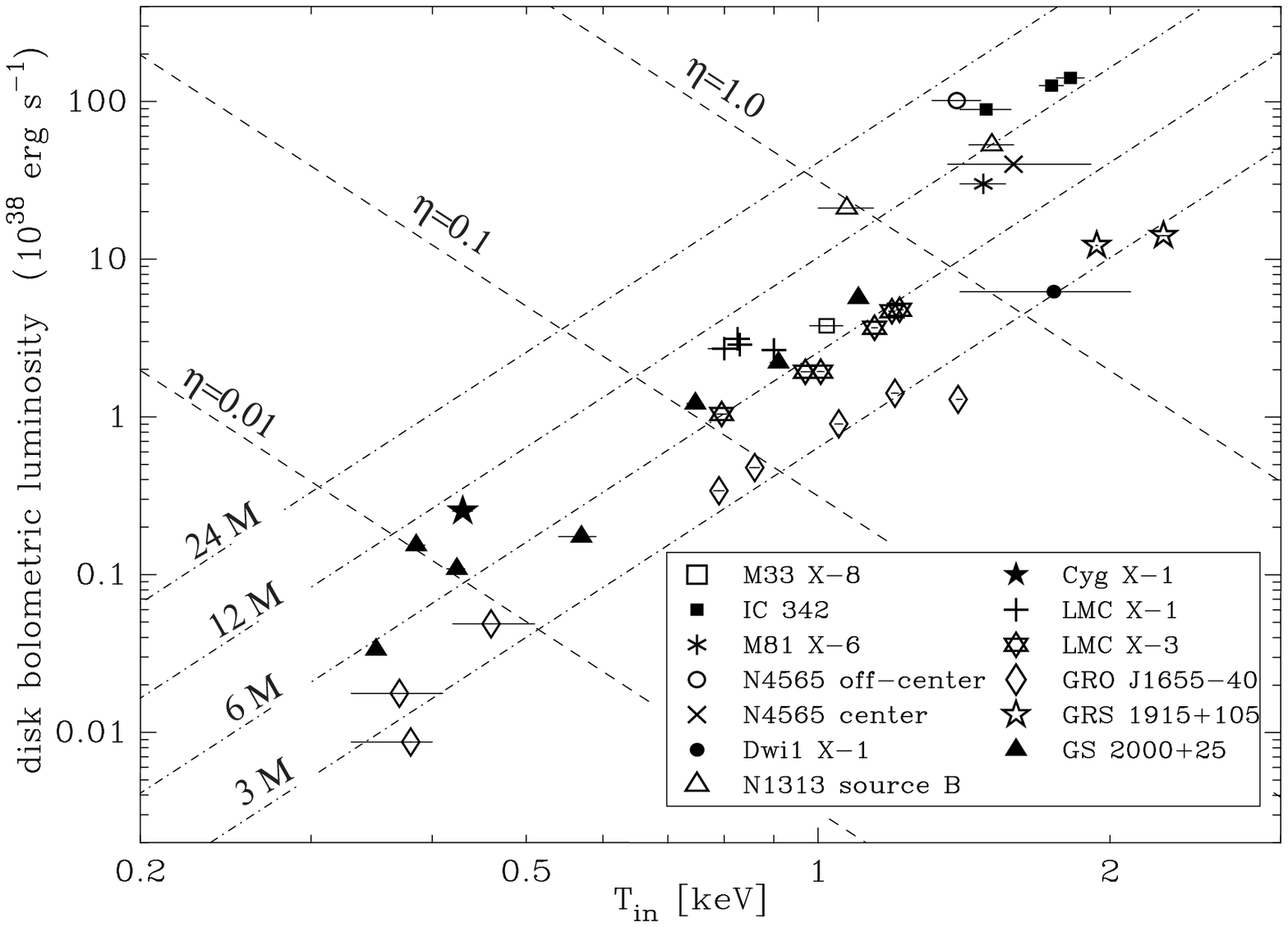}
\figcaption{
  Relation between the bolometric luminosity ($L_{\rm bol}$) 
  and the highest color temperature ($T_{\rm in}$) 
  of optically thick accretion disks around black holes.
  Various symbols indicate measurements with {\it ASCA} and other missions.
  Dash-dotted lines show constant-mass grids,
  calculated via equation (9), assuming the standard accretion disk.
  Dashed lines are those for normalized accretion rates,
  calculated similarly via equation (11).
  In addition to the data described in the text, plotted are 
  two intensity-sorted data points for IC~342 Source 1 (Mizuno 2000),
  three {\it Ginga} data points for LMC~X-1 (Ebisawa 1991), 
  four {\it Ginga} data points for LMC~X-3 (Ebisawa 1991),
  seven {\it RXTE} data points for GRO~J1655-40 (Mendez et al. 1998),
  one {\it RXTE} data points for GRS~1915+105 (Belloni et al. 1997),
  and seven {\it Ginga} data points for GS~2000+25 (Ebisawa 1991).
  Assumed distances refer to Table 1.
  For the ULXs, $i=0$ is assumed.
\label{fig:Aya_plot}
}

\clearpage
\epsscale{0.6}
\plotone{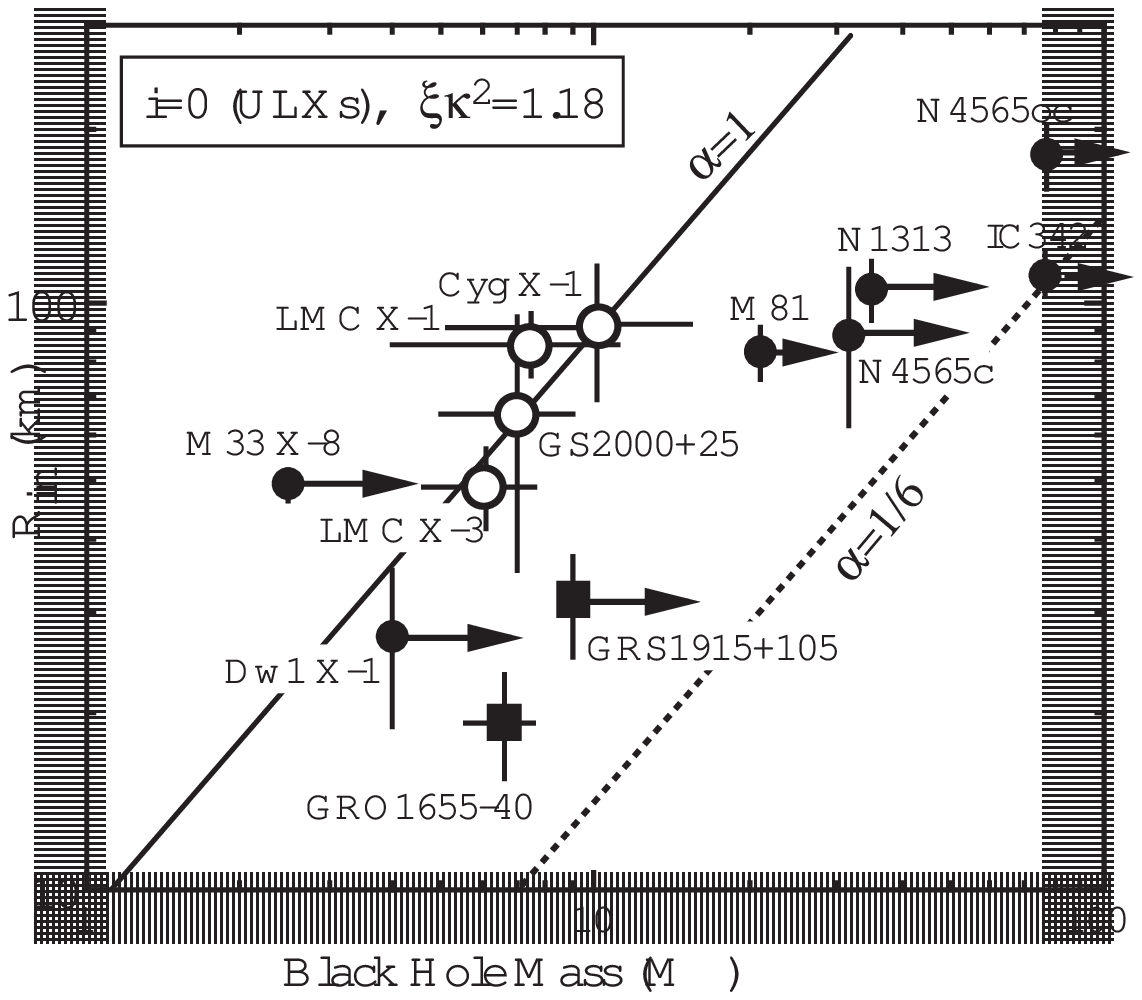}
\figcaption{
 Relation between the innermost radius $R_{\rm in}$ of an accretion disk
 (see Table~1) and the mass of its central black hole.
 The BH mass refers to the optical determination ($M_{\rm opt}$) if available,
 or the Eddington-limit mass $M_{\rm E}$ otherwise
(displayed as lower limits).
 Filled circles, filled squares, and open circles represent ULXs,  
 Galactic jet sources, and other ``ordinary'' BH binaries, respectively.
 The error bars for Galactic and Magellanic objects takes into account
 typical distance uncertainties.
 The point for NGC~1313 refers to the 1993 data.
 The solid and dotted lines show predictions of $\alpha=1$ and $\alpha=1/6$ 
 in equation (7), respectively, both assuming $\xi \kappa^2=1.18$.
\label{fig:MeRin}
}

\clearpage


\begin{table}[hbt]
\caption{Summary of the MCD parameters of ULXs and Galactic/Magellanic BH binaries.$^{a)}$} 
\label{tbl:ULXsummary}
\begin{small}
\begin{center}
\begin{tabular}{clcccccc}
\hline\hline 
&Source &distance &$i \ ^{b)}$&$T_{\rm in} $
                              &$ L_{\rm bol}$& BH Mass $^{c)}$&$R_{\rm in}\ ^{d)}$\\ 
&        &  (pc)    & (deg)  & (keV)  & ($10^{38}$erg/s) &($M_\odot$) &(km)\\
\hline
\multicolumn{3}{l}{ULXs} \\
&M33 X-8 $^{e)}$ &0.72 M & (0)&$1.15\pm 0.03$ & 3.8 & $>2.5$ &$ 49 \pm 3  $   \\
&IC~342 Source~1$ ^{f)}$
                 &4.0 M  & (0) &$1.77\pm 0.05$  & 115 & $>77$ &$113 \pm 10 $   \\
&M81 X-6 $^{g)}$ & 3.6 M & (0) &$1.48\pm 0.08$  &  32 & $>21$ &$ 83 \pm  8 $   \\
&Dwingeloo 1 X-1  $^{g)}$
                 & 3.0 M & (0) &$1.80^{+0.33}_{-0.26}$  
                                            & 6.1 & $>4.1$ &$ 25^{+9}_{-7}$\\  
&NGC~1313 Source~B $^{h)}$
                 & 4.8 M & (0) &$1.47\pm 0.08$  &  57 & $>38$ &$110^{+12}_{-11}$\\  
&               & ---   & (0) &$1.07\pm 0.07$  &  23 & $>15$ &$147 \pm 25$\\  
&NGC~4565 off-center $^{i)}$
                 & 10.4 M & (0) &$1.39\pm 0.08$  & 117 & $>78$ &$185^{+23}_{-17}$\\
&NGC~4565 center  $^{i)}$
            & 10.4 M & (0)&$1.59^{+0.32}_{-0.23}$&  47 & $>31$ &$ 89^{+28}_{-24}$\\
\hline
\multicolumn{4}{l}{Galactic Transients with Jets}\\
&GRO~J1655$-$40 $^{j)}$
      & 3.2 k &70 &$1.394^{+0.014}_{-0.006}$&1.3& $\sim 7$ &$19.4^{+0.1}_{-0.3}$\\
&GRS~1915+105 $^{k)}$
           &12.5 k &70 &$1.937\pm 0.004$ & 12  & $>9$ & $ 30.9^{+0.1}_{-0.2}$  \\
\hline
\multicolumn{5}{l}{Standard Galactic and Magellanic Black-Hole Binaries}\\
&Cyg X-1 $^{l,m)}$ &2.5 k & 30 &$0.43 \pm 0.01$ & 0.25 &$\sim 10$ & $90 \pm 18$\\
&LMC X-1 $^{m)}$   & 55 k & 60 
               &$0.825^{+0.007}_{-0.017}$ & 3.1 & 4--10 & $86^{+2}_{-3}$\\ 
&LMC X-3 $^{m)}$   & 55 k & 66 
               &$0.97 ^{+0.02}_{-0.01}$ & 1.9   & 5.0--7.2 & $49^{+2}_{-3}$  \\
&GS~2000+25 $^{n)}$& 3.0 k& 69 
               &$1.10 \pm 0.01$ & 5.7           & 5--8  & $65.3^{+1.2}_{-1.0}$\\
\hline
\end{tabular} \begin{itemize} 
\setlength{\itemsep}{-1mm}\setlength{\baselineskip}{-5mm}
\item[$^{a)}$]: Based on {\it ASCA} measurements except GS~2000+25. 
\item[$^{b)}$]: The system inclination, 
   assumed to be $i=0^\circ$ if unknown (in parentheses). 
\item[$^{c)}$]: For ULXs and GRS~1915+105, the minimum BH mass 
   to satisfy the Eddington limit is shown. 
   For the other objects, optically estimated BH masses are quoted.  
\item[$^{d)}$]: In all cases, (re-)calculated using equation (5) with 
     $\xi=0.41$ and $\kappa=1.7$.
\item[$^{e)}$]: Taken from Mizuno (2000), which updates Takano et al. (1994).
  The photon index of the power-law hard tail is fixed at 2.2.
\item[$^{f)}$]: Details are given in Table~2.
     Results are consistent with the report by Okada et al. (1998).
\item[$^{g)}$]: Details are given in Table~2. 
\item[$^{h)}$]: The first and second rows refer to the 1993 and 1995 {\it ASCA}
     observations, respectively; see also Table~2 and Petre et al. (1994).
\item[$^{i)}$]: Taken from Mizuno et al. (1999).
\item[$^{j)}$]: The same GIS data as used by Ueda et al. (1998),
    fitted with the MCD model and a power-law. 
    See text \S~4.1, Fig.2a and Fig.2b.
\item[$^{k)}$]: See text \S~4.2 and Fig.2c for detail.
\item[$^{l)}$]: The same GIS data as used by Dotani et al. (1997).
\item[$^{m)}$]: Results from the MCD plus power-law fits.
    See text \S~5.4 for detail.
\item[$^{n)}$]: From Ebisawa (1991), 
    measured with {\it Ginga} on 1989 May 3 near the outburst peak
    and fitted with the MCD model plus a power-law. See text \S~5.5 for detail.
\end{itemize}
\end{center}
\end{small}
\end{table}

\begin{table}[hbt]
\begin{small}
\begin{center}
\caption{Fits to the {\it ASCA} spectra of four ULXs.$^{a)}$} 
\label{tbl:specfits}
\begin{tabular}{llccccc}
\hline\hline 
& Model & Parameter $^{b)}$  & $N_{\rm H} \; ^{c)}$ &$\chi^2/\nu$
                                       &$f_{\rm x}\;^{d)}$ &$f_{\rm bol}\;^{e)}$\\  
\hline
\multicolumn{2}{l}{IC~342 Source~1 (average)}\\
~~& Power-law	     & $1.90 \pm 0.05$      & $9.3 \pm 0.6  $   & 266.5/135 & 15.8 \\
~~& Thermal Brems. & $ 6.1 \pm 0.5 $      & $7.5 \pm 0.4  $   & 183.2/135 & 13.1 \\
~~& MCD $^{f)}$    & $1.77 \pm 0.05$      & $4.7 \pm 0.3  $   & 137.4/135 & 10.7 & 12.2\\
~~& Unsaturated Compton 
          & $1.42 \pm 0.09 / 23 \pm 2$    & $6.4 \pm 0.7  $   & 144.9/134 & 11.5 \\
~~& Broken Power-Law  
     & $1.4 \pm 0.2 / 2.4\pm 0.3 \;^{g)}$ & $6.3 \pm 0.9  $   & 156.3/134 & 12.1 \\
\hline
\multicolumn{3}{l}{M81 X-6}\\
~~& Power-law	     & $1.91 \pm 0.08$      & $4.4 \pm 0.6  $   & 152.8/82  & 5.16 \\
~~& Thermal Brems. & $ 5.4 \pm 0.8 $      & $3.5 \pm 0.4  $   & 113.6/82  & 4.42 \\
~~& MCD $^{f)}$    & $1.48 \pm 0.08$      & $2.1 \pm 0.3  $   &  83.5/82  & 3.69 & 4.10\\
~~& Unsaturated Compton 
               & $1.1 \pm 0.1 / 29 \pm 4$ & $2.5 \pm 0.2  $   &  80.4/81  & 3.94 \\
~~& Broken Power-Law 
 &$1.3\pm 0.2 / 2.8^{+0.7}_{-0.5}\;^{g)}$ & $2.6 \pm 0.4  $   &  85.1/80  & 4.24 \\
\hline
\multicolumn{3}{l}{Dwingeloo 1 X-1}\\
~~& power-law      & $1.87 \pm 0.26 $      &$12.3^{+3.2}_{-2.8}$ & 64.4/59 & 1.51 \\
~~& Thermal Brems. & $7.3^{+5.3}_{-2.4}$   &$10.4^{+2.3}_{-2.1}$ & 63.9/59 & 1.27 \\
~~& MCD $^{f)}$    & $1.80^{+0.33}_{-0.26}$&$7.3^{+2.0}_{-1.6}$  & 64.0/59 & 1.03 & 1.14\\
\hline
\multicolumn{4}{l}{NGC~1313 Source~B, 1993 July} \\
~~&power-law       & $1.99 \pm 0.09$       & $4.0 \pm 0.7  $   & 174.2/130 & 5.43 \\  
~~& Thermal Brems. & $5.1^{+0.9}_{-0.7}$   & $2.6 \pm 0.6  $   & 143.2/130 & 4.60 \\
~~& MCD $^{f)}$    & $1.47 \pm 0.08$       & $0.8 \pm 0.4  $   & 124.5/130 & 3.82 &
4.18\\ \hline
\multicolumn{4}{l}{NGC~1313 Source~B, 1995 November} \\
~~&power-law       & $2.46 \pm 0.13  $     & $4.4 \pm 0.8$     & 110.3/74  & 1.51 \\
~~& Thermal Brems. & $  2.9 \pm 0.4  $     & $2.3 \pm 0.6$     &  91.8/74  & 1.44 \\
~~& MCD $^{f)}$    & $1.07 \pm 0.07  $     & $0.6 \pm 0.4$     &  89.8/74  & 1.39 & 1.66\\
\hline
\end{tabular} 
\end{center}
\begin{itemize} \setlength{\itemsep}{-1mm}\setlength{\baselineskip}{-5mm}
\item[$^{a)}$]: The joint GIS/SIS fits, except the case of M81 X-6
                where only the SIS spectrum is used.
\item[$^{b)}$]:  Photon index for the power-law model, 
     temperature (keV) for the Bremsstrahlung model, 
     $T_{\rm in}$ (keV) for the MCD model, 
     $T_{\rm e}$ (keV)/$\tau_{\rm es}$ for the unsaturated Comptonization model, 
    and low-energy/high-energy photon indices for the broken power-law.
\item[$^{c)}$]: Hydrogen column density for the photoelectric absorption,
     in units of $10^{21}$ cm$^{-2}$.
\item[$^{d)}$]: The 0.5--10 keV flux in $10^{-12}$ ergs s$^{-1}$ cm$^{-2}$, 
     after removing absorption.
\item[$^{e)}$]: The bolometric flux in $10^{-12}$ ergs s$^{-1}$ cm$^{-2}$,
     calculated from the best-fit MCD model.
\item[$^{f)}$]: The values of $R_{\rm in}$ are given in Table~1.
\item[$^{g)}$]: A break energy is at $3.6 \pm 0.5$ keV for IC~342 Source~1,
     and at $3.3 \pm 0.5$ keV for M81 X-6.
\end{itemize}
\end{small}
\end{table}


\begin{thebibliography}{}
\setlength{\itemsep}{-1mm}\setlength{\baselineskip}{-6mm}
\bibitem[]{} Abramowicz, M.A., Czerny, B., Lasota, P., \& Szuszkiewicz, E. 1988, 
             ApJ, 332, 646
\bibitem[]{} Asaoka, I. 1989, PASJ, 41, 763
\bibitem[]{} Beekman, G., Shahbaz, T., Naylor, T., \& Charles, P.A. 1996, MNRAS, 218, L1
\bibitem[]{} Belloni, T., Mendez, M., King, A, van del Klis, M., \& 
	            van Paradijs, J. 1997, ApJ, 479, L145
\bibitem[]{} Beloborodov, A.M. 1998, MNRAS, 297, 739 
\bibitem[]{} Bregman, J. N., Cox, C. V., \& Tomisaka, K. 1993, ApJ, 415, L79
\bibitem[]{} Burke, E.B., Mountain, R.W., Daniels, P.J., Cooper, M.J, \& Dolat, V.S.
             1994, IEEE Trans. Nuc. Sci. 41, 375
\bibitem[]{} Callanan, P., Garcia, M., Filippenko, A., McLean, I., 
             \& Teplitz, H. 1996, ApJ, 470, L57
\bibitem[]{} Chen, W., Cui, W., \& Zhang, S.N. 1997, AAS Meeting, 191, 59.07
\bibitem[]{} Colbert, E.J.M., Petre, R., Schlegel, E.M., \& Ryder, S.D. 1995, 
             ApJ, 446, 177
\bibitem[]{} Colbert, E.J.M., \& Mushotzky, R.F. 1999, ApJ, 519, 89
\bibitem[]{} Cowley, A.P. 1992, ARA\&A, 30, 287
\bibitem[]{} Cowley, A.P., Schmidtke, P.C., Anderson, A.L., McGarth, T.K. 1995, 
             PASP 107, 145
\bibitem[]{} Dotani T., et al. 1997, ApJ 485, L87 
\bibitem[]{} Dubas, G., Charles, P., Long, K.S., \& Hakara, P.J. 1997, ApJ, 490, L47
\bibitem[]{} Ebisawa, K., 1991, PhD Thesis, University of Tokyo
\bibitem[]{} Ebisawa, K., 1997, X-Ray Imaging and Spectroscopy of Cosmic Hot Plasmas,
             eds. F. Makino and K. Mitsuda (Universal Academy Press; Tokyo), p427
\bibitem[]{} Ebisawa, K., Mitsuda, K., \& Hanawa, T. 1991, ApJ, 367, 213
\bibitem[]{} Ebisawa, K., Makino, F., Mitsuda, K., Belloni, T., Cowley, A., 
             Schmidke, P., \& Treves, A. 1993, ApJ, 403, 684
\bibitem[]{} Esin, A., McClintock, J.E., \& Narayan, R. 1997, ApJ, 489, 865
\bibitem[]{} Fabbiano, G. 1988a, ApJ, 325, 544 
\bibitem[]{} Fabbiano, G. 1988b, ARA\&A, 27, 87
\bibitem[]{} Fabbiano, G., \& Trinchieri, G. 1987, ApJ, 315, 46
\bibitem[]{} Fabbiano, G., Trinchieri, G., \& van Speybroeck 1987, ApJ, 316, 127 
\bibitem[]{} Filippenko, A., Matheson, T., \& Barth, A.A. 1995, ApJ, 455, L139
\bibitem[]{} Freedman, W. L., et al. 1994, ApJ 427, 628
\bibitem[]{} Fryer, C.L. 1999, ApJ, 522, 413
\bibitem[]{} Gottwald, M., Pietsch, W., \& Hasinger, G. 1987, A\&A 175, 45  
\bibitem[]{} Hanawa, T. 1989, ApJ, 341, 948 
\bibitem[]{} Herreo, A., Kudritzki, R.P., Gabler, R., Vilchez, J. M., 
             \& Gabler, A. 1995, A\&A, 297, 556
\bibitem[]{} Hjellming, R.M., \& Rupen, M.P. 1995, Nature, 375, 464
\bibitem[]{} Ishisaki, Y., et al. 1996, PASJ, 48, 237
\bibitem[]{} Iyomoto, N. 1999, PhD Thesis, University of Tokyo
\bibitem[]{} Kohmura, Y., 1994, PhD Thesis, University of Tokyo
\bibitem[]{} Kohmura, Y., et al. 1994, PASJ, 46, L157
\bibitem[]{} Kotani, T., et al. 2000, Adv. Sp. Res., in press
\bibitem[]{} Krabbe, A., et al. 1995, ApJ, 447, L95.
\bibitem[]{} Kubota, A., Tanaka, Y., Makishima, K., Ueda, Y., Dotani, T., Inoue, H.,
            \& Yamaoka, K. 1998, PASJ, 50, 667
\bibitem[]{} Kuiper, L., van Paradijs, J., \& van der Klis, M. 1988, A\&A, 203, 79
\bibitem[]{} Lauer, T., Faber, S., Ajhar, E., Grillmair, C., \& Scowen, P.
             1998, AJ, 116, 2263
\bibitem[]{} Long, K. S. 1982, Adv.Sp.Res. 2, 177
\bibitem[]{} Long, K. S., D'Odorico, S., Charles, P. A., \& Dopita, M. A. 1981, 
             APJ, 246, L61 
\bibitem[]{} Long, K. S., Charles, P. S., Blair, W. P., \& Gordon, S. M. 1996, 
             ApJ, 466, 750
\bibitem[]{} Makishima, K. 1994, New Horizon of X-Ray Astronomy, 
             eds. F. Makino and T. Ohashi (Universal Academy Press, Tokoy), p171
\bibitem[]{} Makishima, K., et al. 1986, ApJ 308, 635 
\bibitem[]{} Makishima, K., et al. 1989, PASJ, 41, 697
\bibitem[]{} Makishima, K., et al. 1996, PASJ 48, 171 
\bibitem[]{} Marston, A.P., Elmegreen, D., Elmegreen, B., Forman, W., Jones, C.,
            \& Flanagan, K. 1995, ApJ, 438, 663
\bibitem[]{} Mendez, M., Belloni, T., \& van der Klis, M. 1998, ApJ, 499, L187
\bibitem[]{} Mineshige, S., Hirano, A., Kitamoto, S., \& Yamada, T. 1994, ApJ, 426, 308
\bibitem[]{} Mirabel, I.F., \& Rodriguez, L.F. 1994, Nature, 371, 46
\bibitem[]{} Mitsuda K., et al. 1984, PASJ 36, 741 
\bibitem[]{} Mizuno, T., 2000, PhD Thesis, University of Tokyo
\bibitem[]{} Mizuno, T., Ohnishi, T., Kubota, A., Makishima, K., \& Tashiro, M.
             PASJ, 51, 663
\bibitem[]{} Narayan, R., \& Yi, I, 1995, ApJ, 452, 710
\bibitem[]{} Ohashi T., et al. 1996, PASJ 48, 157 
\bibitem[]{} Okada, K., Dotani, T., Makishima, K., Mitsuda, K., \& Mihara, T. 1998,
             PASJ, 50, 25
\bibitem[]{} Orosz, J.A., \& Bailyn, C. D. 1997, ApJ, 477, 876
\bibitem[]{} Petre, R., Okada, K., Mihara, T., Makishima, K., \& Colbert, E.J.M.,
             1994, PASJ, 46, L115
\bibitem[]{} Raymond, J.C., \& Smith, B.W. 1977, ApJS, 35, 419
\bibitem[]{} Read, A.M., Ponman, T.J., \& Strickland, D.K. 1997, MNRAS, 286, 626 
\bibitem[]{} Reynolds, C.S, \& Begelman, M.C. 1997, ApJ, 488, 109
\bibitem[]{} Reynolds, C.S., Loan, A.J., Fabian, A.C., Makishima, K., Brandt, W.N.,
             \& Mizuno, T. 1997, MNRAS, 286, 349 
\bibitem[]{} Schlegel, E.M. 1994, ApJ, 424, L99
\bibitem[]{} Shahbaz, T., van del Hooft, F., Casares, J., Charles, P.A., \& 
	            van Paradijs, J. 1999, MNRAS, 306, 89
\bibitem[]{} Shakura, N.I., \& Sunyaev, R.A. 1973, A\&A, 24, 337
\bibitem[]{} Shimura, T., \& Takahara, F. 1995, ApJ, 445, 780
\bibitem[]{} Sunyaev, R.A., \& Titarchuk, L.G. 1980, A\&A, 86, 121
\bibitem[]{} Supper, R., et al. 1997, A\&A 317, 328
\bibitem[]{} Takano, M., Mitsuda, K., Fukazawa, Y., \& Nagase, F. 1994, ApJ, 436, L47
\bibitem[]{} Tanaka, Y. 1997, in Accretion Disk - New Aspects, 
        ed E. Meyer-Hofmeister, H. Spruit, Lecture Note in Physics Vol. 487
\bibitem[]{} Tanaka, Y., Inoue, H., Holt, S.S. 1994, PASJ, 46, L37 
\bibitem[]{} Tanaka, Y., \& Lewin, W.H.G. 1995, in X-ray Binaries, 
		   eds. W.H.G. Lewin, J. van paradijs, and W.P.J. van den Heuvel 
    (Cambridge University Press, Cambridge), p126
\bibitem[]{} Tanaka, Y., \& Shibazaki, N. 1996, ARA\&A, 34, 607
\bibitem[]{} Trinchieri, G., Fabbiano, G., \& Peres, G. 1988, ApJ, 325, 531
\bibitem[]{} Tsunemi, H., Kotamoto, S., Okamura, S., Roussel-Dupr\'{e}, D.
   1989, ApJ, 337, L81
\bibitem[]{} Ueda, Y., Inoue, H., Tanaka, Y., Ebisawa, K., Nagase, F., Kotani, T., 
  \& Gehrels, N. 1998, ApJ, 492, 782
\bibitem[]{} Uno, S., 1997, PhD Thesis, Gakushuin University
\bibitem[]{} van Speybroeck, L., Epstein, A., Forman, W., Giacconi, R., Jones, C.,
     Lillter, W., \& Smarr, L. 1979, ApJ, 234, L45
\bibitem[]{} Watarai, K., Fukue, J., Takeuchi, M., \& Mineshige, S. 1999,
      PASJ, submitted
\bibitem[]{} White, N.E., \& Marshall, F.E. 1984, ApJ, 281, 354
\bibitem[]{} Yamashita A., et al. 1997, IEEE Trans. Nucl. Sci., 44, 847 
\bibitem[]{} Zhang, S.N., Cui, W., \& Chen, W. 1997, ApJ, 482, L155
\bibitem[]{} Zhang, S.N., et al. 1997b, ApJ, 479, 381
\end{thebibliography}
\end{document}